# Spin-lattice relaxation phenomena in manganite $La_{0.7}Sr_{0.3}MnO_3$ thin films

M. C. Weber * and B. Hillebrands

Fachbereich Physik und Forschungsschwerpunkt MINAS, Technische Universität

Kaiserslautern, Erwin-Schrödinger-Str. 56, 67663 Kaiserslautern, Germany

V. Moshnyaga and K. Samwer

I. Physikalisches Institut, Universität Göttingen, Friedrich-Hund-Platz 1,

37077 Göttingen, Germany

* Electronic mail: mweber@physik.uni-kl.de

Time-resolved magneto-optics was used to study spin-lattice relaxation dynamics in thin epitaxial La$_{0.7}$Sr$_{0.3}$MnO$_3$ films. Two distinct recovery regimes of the ferromagnetic order can be resolved upon photoexcitation, which manifest themselves by two different relaxation times. A pump pulse energy independent spin-lattice relaxation time can be deduced. Due to a weak spin-orbit coupling in manganites this spin-lattice relaxation time is much longer than in ferromagnetic metals. Heat flow into the substrate sets the ultimate recovery speed of the ferromagnetic order and allows for a determination of heat diffusion properties of manganite films.

PACS number(s): 78.47.+p , 75.70.-i , 75.40.Gb

Significant research effort has been focused on ultrafast magnetization and demagnetization dynamics in the recent years. An all-optical technique employing short laser pulses may be of interest, as field pulse assisted magnetization switching cannot be faster than a magnetization precession cycle[1]. However, a special demagnetization mechanism is required - an optical pulse can barely change the magnetization directly. Ultrafast magnetization dynamics has been studied in ferromagnetic[2-4] and various other material systems[5,6]. On the other hand, colossal magnetoresistive manganites are promising candidates for ultrafast spin control due to coupled spin, charge, orbital and lattice degrees of freedom[7].

Perovskite manganites exhibit an insulator to metal transition usually related to a para-ferromagnetic phase transition at the Curie temperature. The origin of ferromagnetism is double-exchange interaction[8] yielding a strong correlation between magnetization and charge transport properties. Photoinduced effects on the picosecond timescale for the ferromagnetic phase of manganites have been previously studied by conventional pump-probe spectroscopy[9-11]. The first report on laser-induced demagnetization was presented by Matsuda *et al.*[9]. A rapid change of the photoinduced absorption within 1-2 ps and a following gradual change up to about 200 ps was found. The ultrafast component is attributed to electron-phonon thermalization, whereas the subsequent slower component with a temperature dependent time constant to spin and lattice thermalization. However, the evaluated time scales for the demagnetization dynamics were based on changes in optical absorption. Thus, spin channels, such as spin-lattice relaxation, have not been discerned in the absorption changes.

In the present work, magnetization dynamics in epitaxial $La_{0.7}Sr_{0.3}MnO_3$ (LSMO) films was investigated by time-resolved magneto-optics where the Kerr rotation was employed as a reliable probe of magnetization[12]. Special emphasis was put on the determination of the spin-lattice relaxation time, a key quantity to answer the question about the fundamental speed limit of spin manipulation in this material. Upon arrival of a short laser pulse a collapse of the ferromagnetic order takes place accompanied by spin-lattice thermalization. The deduced spin-lattice time is considerably longer than in normal ferromagnetic metals. Heat diffusion effects, in particular heat flow into the substrate, set the ultimate speed limit of recovery of the magnetic order and allow for the determination of heat diffusion properties of thin manganite films.

Ferromagnetic perovskite LSMO films were grown on MgO(100) substrates by a metalorganic aerosol deposition technique described elsewhere[13,14]. X-ray diffraction analysis reveals a "cube-on-cube" epitaxy with a pseudocubic c-lattice constant c=0.388 nm. This value is in agreement with the bulk lattice parameter for LSMO, thus, evidencing the absence of mechanical stress in the grown films. The time-resolved measurements have been performed employing a standard all-optical pump-probe setup. An amplified picosecond mode-locked $Nd:YVO_4$ pulsed laser source[15] in combination with second harmonic generation (532 nm) delivers laser pulses of a 8 ps duration. A pulse picker reduces the repetition rate of the system to 1 MHz in order to avoid any dc lattice heating effects. The beam is then divided into an intense pump and a weak probe beam. The pump beam is focused to about 30 μm onto the sample yielding a maximum power density of < 1.2 $GW/cm^2$ for the presented results. The time delayed probe beam is used to detect the induced magnetic changes by means of longitudinal Kerr magnetometry probing the Kerr rotation $\theta_K$. In order to investigate magnetization dynamics in the time domain, a quasi-static hysteresis loop is sensed by a probe pulse with a fixed time delay to the excitation pulse. It then reflects the magnetic parameters, such as the spin temperature present for a given time delay.

First, the absolute longitudinal moment of a 76 nm thick LSMO film as a function of temperature at a field of 100 Oe was measured in a SQUID magnetometer with the field aligned parallel to the substrate. Figure 1 summarizes the SQUID measurements for a temperature range from 4 to 375 K. The LSMO film exhibits a ferromagnetic-paramagnetic phase transition at a Curie temperature of $T_C$ = 357 K. The inset of Fig. 1 shows the temperature dependence of the coercivity in the temperature range of interest. A coercivity of about 11 Oe at room temperature indicates a good epitaxial film quality and the absence of structural pinholes.

Now, we address the question whether it is possible to induce a spin-lattice nonequilibrium close to the ferromagnetic-paramagnetic phase transition in thin LSMO via a short picosecond photoexcitation. Using the above described time-resolved measurement scheme we deduced the time evolution of the Kerr amplitude at zero applied field of the recorded transient hysteresis loops (see inset Fig. 2a) for pump-probe delay times $\Delta t$ up to 4000 ps. The time evolution of the Kerr amplitude at zero applied field will be further denoted as the time evolution of the remanence, thus, directly reflecting the ferromagnetic order as well as the spin temperature of the LSMO film. The time evolution of the remanence exemplarily presented in Fig. 2 for a pump pulse energy of 27.5 nJ clearly indicates a collapse of the magnetic order within the excitation duration, followed by a spin and lattice thermalization based relaxation back to its initial value. The overall equilibrium relaxation dynamics can be well described by the time evolution of the lattice temperature[16]

$$\Delta T_{latt}(t) = \Delta T_0 \left(1 - \exp(-t/\tau_D)\right) \tag{1}$$

which yields a time constant of $\tau_D = 590 \pm 15\,\text{ps}$ for the diffusive processes for the example shown in Fig. 2 (see dash-dotted line in Fig. 2a). If one zooms around zero delay one can clearly see a strong deviation from the diffusive background after application of the pump pulse. A second order exponential recovery fit to the data (see solid line in Fig. 2b) reveals two different relaxation times $\tau_1 = 130 \pm 10\,\text{ps}$ and $\tau_2 = 550 \pm 10\,\text{ps}$. Moreover, by subtracting the diffusive background (Eq. (1)) from the measured data we can clearly resolve the transition between these two distinct relaxation regimes of the ferromagnetic order at $\tau_{sl} = 250 \pm 25\,\text{ps}$ (see inset of Fig. 2b).

The existence of two distinct relaxation regimes during the measured demagnetization and remagnetization process of the ferromagnetic order can be understood considering a strong spin-lattice nonequilibrium upon photoexcitation. The time scale of hot electron relaxation dynamics is below the time resolution of our setup. Hence, thermalization of the spin and phonon systems lasting from the low picosecond up to the nanosecond time scale dominate the measured time evolution of the remanence. An increase in spin temperature upon excitation is followed by a spin-lattice nonequilibrium and subsequently by spin-lattice thermalization mediated via spin-orbit coupling. However, spin-orbit coupling in perovskite manganites is small[12], which is consistent with the negligible magnetocrystalline anisotropy found in the investigated LSMO thin film. Thus, the weak spin-orbit coupling in LSMO accounts for the deviation from the diffusive background (see Fig. 2b) which can be attributed to spin-lattice relaxation dynamics[12]. The data analysis yields a spin-lattice relaxation time window from $\tau_1 = 130 \pm 10$ ps up to $\tau_{sl} = 250 \pm 25$ ps – an upper limit for the relaxation process. This spin-lattice time window is about a factor of 2-3 larger than values reported from Vaterlaus *et al.*[17] and Guarisco *et al.*[18] for typical ferromagnets and at least one order of magnitude larger than those reported for Ni[4], where spin-orbit coupling is large compared to LSMO.

For pump-probe delay times larger than $\tau_{sl}$, where spin and lattice temperatures are in equilibrium, heat diffusion processes dominate the recovery dynamics of the ferromagnetic order of the LSMO film ($\tau_D \simeq \tau_2$). Comparing the time evolution of the spin temperature with the heat diffusion equation[16] one can determine consistently both the spin-lattice time and the heat diffusion properties of the LSMO film, which determine the ultimate speed of recovery of the ferromagnetic order.

Taking into account the temperature dependence of the longitudinal magnetization $M(T)$ from Fig. 1 and the time evolution of the Kerr amplitude at zero field from Fig. 2 one can calibrate the spin temperature evolution. For the given longitudinal Kerr geometry one can neglect any out-of-plane magnetization component contribution to the measured Kerr signal which then reads $\theta_K \simeq \theta_K^{long} \cdot m_y$, where $\theta_K^{long}$ denotes the static Kerr amplitude and $m_y$ the normalized longitudinal magnetization component. This allows for a direct comparison of the reduction of the longitudinal magnetic moment with the elevated spin temperature and the measured transient Kerr rotation correspondingly. The increase in spin temperature $\Delta T$ can be normalized to the maximum spin temperature increase upon excitation ($\Delta T_{max}$ = 52 K). Assuming that one dimensional heat diffusion dominates one can solve the heat flow equation. Moreover, assuming that the pump pulse leads to a Gaussian temperature profile at $\Delta t = \tau_{sl}$ the spin temperature increase[16]

$$\Delta T(\Delta t \geq \tau_{sl}) = \frac{\lambda_0 \cdot \Delta T_{max}}{\sqrt{\pi \cdot D \cdot t + \lambda_0^2}} \quad (2)$$

can be compared to the experimental spin temperature evolution with a thermal diffusion length $\lambda_0$ and the diffusivity $D$. Figure 3a shows the spin temperature evolution together with a fit to Eq. (2), which yields $\lambda_0$ = 25 nm and $D = 3.85(5) \cdot 10^{-6} \, m^2/s$. Considering a composition weighted specific heat of $C \simeq 1.2 \cdot 10^6 \, J/m^3 K$ one can deduce a heat conductivity perpendicular to the film surface of $K = 4.7(1) W/mK$, which agrees well with recently published values[19]. However, it is considerably lower than in typical metals. Since Eq. (2) only fits for pump-probe delay times larger than $250 \pm 50$ ps, this value corresponds to the spin-lattice relaxation time which is consistent with the results found for the time evolution of the ferromagnetic order. Again, a strong deviation from the diffusion dominated relaxation can be identified for delays less than 250 ps. The deduced spin-lattice time window was found to be independent of the energy of the applied pump pulses (see Fig. 3b) which suggests the fundamental nature of the observed long spin-lattice relaxation time in LSMO thin films. However, the time constant of the heat flow process $\tau_D \simeq \tau_2$ depends strongly on the applied pulse energy.

In summary, we have observed spin-lattice relaxation phenomena in thin epitaxial LSMO manganites by time-resolved magneto-optics. A nearly complete breakdown of the ferromagnetic order could be induced and observed on the picosecond time scale. A spin-lattice relaxation time window from 130 ps up to about 250 ps was deduced which is considerably slower compared to normal ferromagnets. The origin lies in the weak spin-orbit coupling present in perovskite manganites. A-site doping of thin LSMO as well as tuning of mechanical strain, thereby modifying the spin-orbit coupling seem to be relevant to control the fundamental spin-lattice relaxation time of an excited LSMO spin state. A clear separation of spin-lattice relaxation dynamics and heat flow within the relaxation dynamics of the magnetic order upon photoexcitation could be observed. Two distinct relaxation times manifest themselves which makes the determination of spin-lattice thermalization and moreover of heat diffusion properties of LSMO films feasible.

The authors acknowledge support by the European Communities Human Potential program HPRN-CT-2002-00318 ULTRASWITCH and by the DFG via SFB 602, TP A2. M.C.W. acknowledges support by the DFG Graduiertenkolleg 792.

## References

[1] M. Bauer, R. Lopusnik, J. Fassbender, and B. Hillebrands, Appl. Phys. Lett. **76**, 2758 (2000).

[2] E. Beaurepaire, J.-C. Merle, A. Daunois, and J.-Y. Bigot, Phys. Rev. Lett. **76**, 4250 (1996).

[3] J. Hohlfeld, J. Güdde, U. Konrad, O. Dühr, H. Korn, and E. Matthias, Appl. Phys. B **68**, 505 (1999).

[4] B. Koopmans, M. van Kampen, J.T. Kohlhepp, and W.J.M. de Jonge, Phys. Rev. Lett. **85**, 844 (2000).

[5] J.M. Kikkawa and D.D. Awschalom, Phys. Rev. Lett. **80**, 4313 (1998).

[6] T. Kise, T. Ogasawara, M. Ashida, Y. Tomioka, and M. Kuwata-Gonokami, Phys. Rev. Lett. **85**, 1986 (200).

[7] Y. Tokura and Y. Tomioka, J. Magn. Magn. Mater. **200**, 1 (1999).

[8] C. Zener, Phys. Rev. **81**, 440 (1951); P.W. Anderson and H. Hasegawa, Phys. Rev. **100**, 675 (1955); P.-G. de Gennes, Phys. Rev. **118**, 141 (1960).

[9] K. Matsuda, A. Machida, Y. Moritomo, and A. Nakamura, Phys. Rev. B **58**, 4203 (1998).

[10] A.I. Lobad, R.D. Averitt, C. Kwon, and A.J. Taylor, Appl. Phys. Lett. **77**, 4025 (2000).

[11] M. Sasaki, G.R. Wu, W.X. Gao, H. Negishi, M. Inoue, and G.C. Xiong, Phys. Rev. B **59**, 12425 (1999).

[12] T. Ogasawara, M. Matsubura, Y. Tomioka, M. Kutawa-Gonokami, H. Okamoto, and Y. Tokura, Phys. Rev. B **68**, 180407(R) (2003).

[13] S.A. Köster, V. Moshnyaga, K. Samwer, O.I. Lebedev, G. van Tendeloo, O. Shapoval, and A. Belenchuk, Appl. Phys. Lett. **81**, 1648 (2002).

[14] V. Moshnyaga, I. Khoroshun, A. Sidorenko, P. Petrenko, A. Weidinger, M. Zeiler, B. Rauschenbach, R. Tidecks, and K. Samwer, Appl. Phys. Lett. **74**, 2842 (1999).

[15] J. Kleinbauer, R. Knappe, and R. Wallenstein, in *Femtosecond Technology for Technical and Medical Applications*, edited by F. Dausinger, F. Lichtner, and H. Lubatschowski (Springer, Berlin, 2004), Topics Appl. Phys. **96**, 17 (2004).

[16] J.H. Bechtel, J. Appl. Phys. **43**, 1585 (1975).

[17] A. Vaterlaus, T. Beutler, and F. Meier, Phys. Rev. Lett. **67**, 3314 (1991).

[18] D. Guarisco, R. Burgermeister, C. Stamm, and F. Meier, Appl. Phys. Lett. **68**, 1729 (1996).

[19] A. Salazar, A. Oleaga, and D. Prabhakaran, Int. J. Thermophys. **25**, 1269 (2004).

**Fig. 1** Temperature dependence of the longitudinal magnetic moment of the 76 nm thin LSMO film, measured by SQUID in a field of $H = 100$ Oe. The dashed lines indicate room temperature and the room temperature moment, respectively. The inset shows the temperature dependence of the coercivity in the range of interest.

**Fig. 2 (a)** Time evolution of the remanence for a pump pulse energy of 27.5 nJ. The dash-dotted line highlights the diffusive relaxation regime with a time constant of $\tau_D = 590$ ps. The inset shows three hysteresis loops for a negative (< 0), positive (>0) and zero pump-probe delay. **(b)** Spin-lattice thermalization regime within the recovery process of the ferromagnetic order. The solid line represents a 2$^{nd}$ order exponential recovery ($\tau_1 = 130$ ps and $\tau_2 = 550$ ps). The inset shows a strong deviation from the diffusive background for $\Delta t < 250$ ps.

**Fig. 3 (a)** Time evolution of the spin temperature upon photoexcitation with a fit to Eq. (2) yielding a spin-lattice relaxation time of about $\tau_{sl} \leq 250$ ps. **(b)** Pump pulse energy dependence of the deduced spin-lattice relaxation time window ($\tau_1$, $\tau_{sl}$), which stays constant within the experimental error.

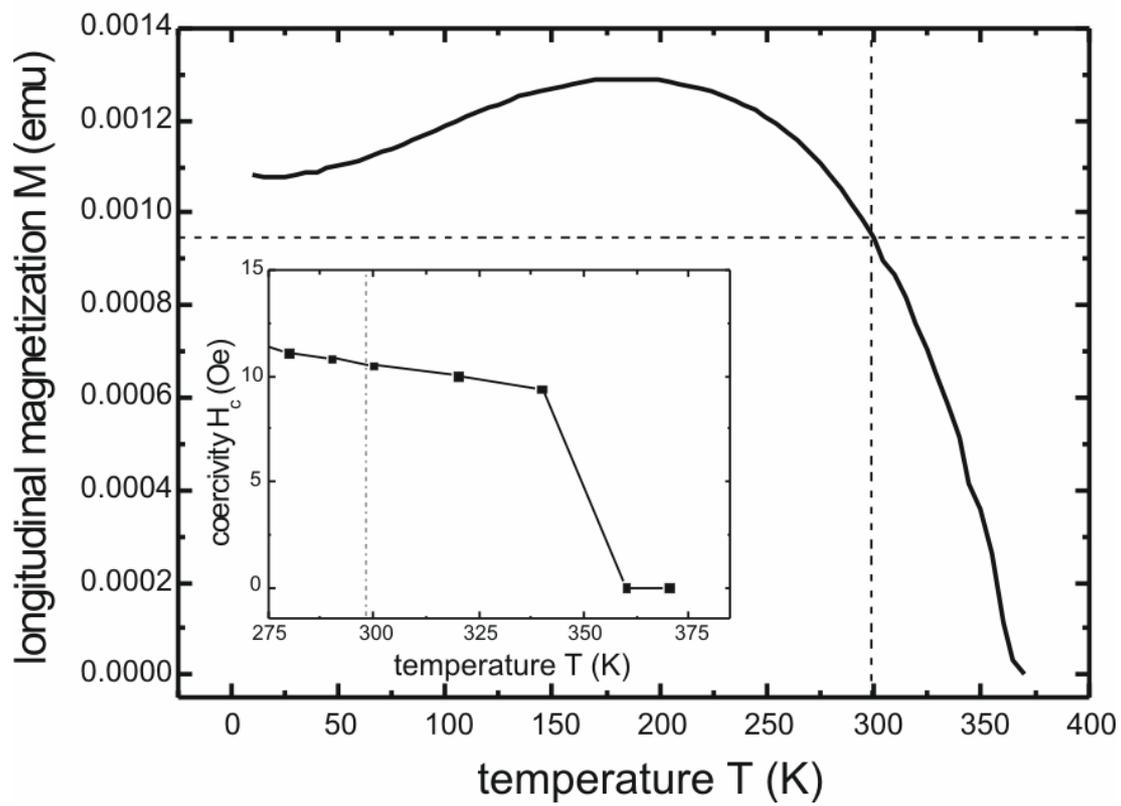

**Fig. 1**

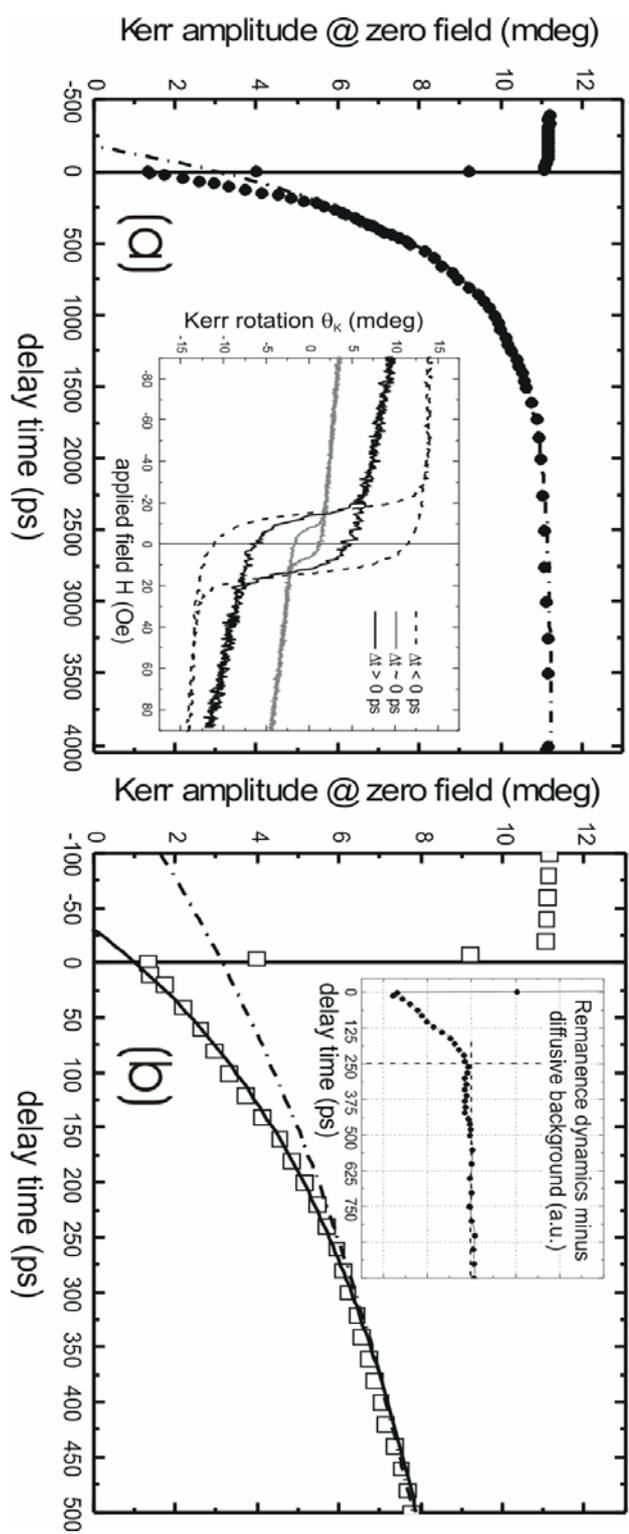

**Fig. 2**

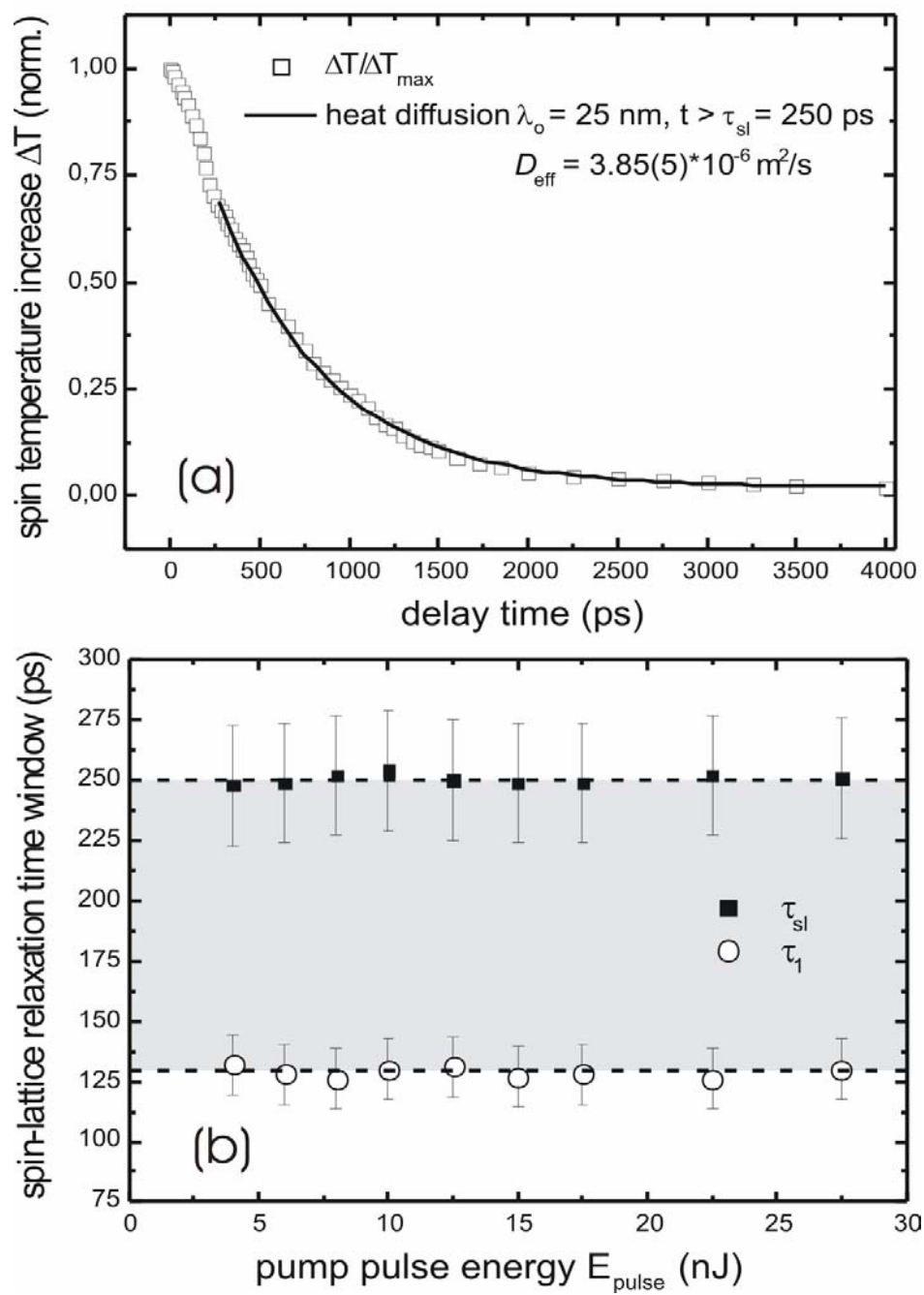

**Fig. 3**